\documentclass[prb,twocolumn,showpacs,superscriptaddress,preprintnumbers,amsmath,amssymb]{revtex4-1}
\usepackage{graphicx} \usepackage{dcolumn} \usepackage{bm} \usepackage{amsmath} \usepackage{color}
\usepackage{hyperref} \usepackage[utf8]{inputenc} \usepackage[T1]{fontenc}

\begin{document}


\title{Coupled spin-lattice fluctuations in a compound with orbital degrees of freedom: \\
the Cr based dimer system Sr$_3$Cr$_2$O$_8$}

\author{Dirk Wulferding} \author{Peter Lemmens} \affiliation{Institute for Condensed Matter
Physics, Technical University of Braunschweig, D-38106 Braunschweig, Germany}

\author{Kwang-Yong Choi} \affiliation{Department of Physics, Chung-Ang University, 221
Huksuk-Dong, Dongjak-Gu, Seoul 156-756, Republic of Korea}

\author{Vladimir Gnezdilov} \affiliation{B.I. Verkin Inst. for Low Temperature Physics and
Engineering, NASU, 61103 Kharkov, Ukraine}

\author{Joachim Deisenhofer} \affiliation{Experimental Physics V, University of Augsburg, D-86135 Augsburg, Germany}

\author{Diana Quintero-Castro}
\affiliation{Helmholtz Zentrum Berlin f\"{u}r Materialien und Energie, D-14109 Berlin, Germany}
\affiliation{Institut f\"{u}r Festk\"{o}rperphysik, Technische Universit\"{a}t Berlin, D-10623 Berlin, Germany}

\author{A.T.M. Nazmul Islam}
\affiliation{Helmholtz Zentrum Berlin f\"{u}r Materialien und Energie, D-14109 Berlin, Germany}

\author{Bella Lake}
\affiliation{Helmholtz Zentrum Berlin f\"{u}r Materialien und Energie, D-14109 Berlin, Germany}
\affiliation{Institut f\"{u}r Festk\"{o}rperphysik, Technische Universit\"{a}t Berlin, D-10623 Berlin, Germany}

\date{\today}

\begin{abstract} We report on an extended fluctuation regime in the spin dimer system
Sr$_3$Cr$_2$O$_8$ based on anomalies in Raman active phonons and magnetic scattering. The
compound has two characteristic temperatures, $T_S = 275$ K, related to a Jahn-Teller transition with structural distortions and orbital ordering and a second, $T^* \approx 150$ K, which is due to further changes in the orbital sector. Below $T_S$ quasielastic scattering marks strong fluctuations and in addition phonon anomalies are observed. For temperatures below $T^*$ we observe an exponential decrease of one phonon linewidth and determine a gap of the orbital excitations. At low temperatures the observation of two- and three-magnon scattering allows the determination of the spin excitation gap.
\end{abstract}


\maketitle

\section{Introduction} The interplay of electronic correlation effects with other interactions in
transition metal oxides can lead to a wealth of interesting phenomena such as colossal
magneto-resistance, exotic superconductivity and magnetism including rich phase diagrams
\cite{tokura}. The Jahn-Teller instability is a well known example of coupled orbital --
structural instabilities where the orbital degeneracies of the $3d$ electron states can be lifted
by distortions of the surrounding oxygen coordinations. These variations in the occupation of
orbital states and related lattice distortions may have a collective, long range or a local,
fluctuating character. Interesting from fundamental and applied points of view
are cases where fluctuation regimes cross over from glass like to quantum disordered states. As
examples pseudo-cubic manganites and titanates (LaTiO$_3$, YTiO$_3$) are mentioned. In these
compounds quantum disordered states based on orbital degrees of freedom are discussed
\cite{khaliullin01, sawada}.

It is presently a state of debate how the excitation spectrum of such systems can be characterized
and whether collective orbital/electronic excitations or modes of mixed phonon-orbital origin
exist. This question gains complexity from the fact that in purely electronic models quantum
orbital liquids with entangled orbital states can be stabilized by electronic correlations. The
relevance of such models for real materials is an interesting issue as in addition to electron-phonon
interaction, thermal fluctuations and doping have to be considered. Investigations have focused so far on cubic manganites ($R,A$MnO$_3$, Mn$^{3+}$, $3d^4$) \cite{saitoh, choi-lsmo}, titanates (LaTiO$_3$, Ti$^{4+}$, $3d^1$) \cite{ulrich, konstantinovic}, vanadates ($R$VO$_3$, V$^{3+}$, $3d^1$) \cite{miyasaka} and iridates (CuIr$_2$S$_4$, Ir$^{3+}$/Ir$^{4+}$, $5d^6$/$5d^5$) \cite{radaelli}. Only recently also compounds based on chromium (Cr$^{5+}$, $3d^1$) in
tetrahedral coordination have been discussed. Besides other techniques, Raman scattering has been used to investigate electronic as well as lattice degrees of freedom.

In the compound Sr$_3$Cr$_2$O$_8$ the transition metal ion Cr$^{5+}$ is coordinated by four oxygen ions in a tetrahedral environment. At room temperature the structure is hexagonal ($R\overline{3}m$) and
bilayers of CrO$_4$ tetrahedra form a frustrated arrangement of two degenerate $3d$ $e$ levels
$d_{3z^2-r^2}$ and $d_{x^2-y^2}$. For temperatures below $T_S = 275$ K crystallographic
distortions lead to monoclinic structure ($C2/c$) and a splitting of the $e$ levels with primary
occupation and antiferro-orbital ordering of the lower-lying $d_{3z^2-r^2}$ orbitals \cite{chapon, radtke}. Assuming static orbital order, the low energy magnetism is based
on Heisenberg exchange coupled spin $s = 1/2$ dimers of the pairwise arranged
tetrahedra. This leads to a spin gap $\Delta$ of the magnetic excitations.

The magnetic properties follow this scenario pretty well: The magnetic susceptibility has a weak anisotropy, shows a maximum at $T_{max} = 38$ K and a rapid decrease towards lower temperatures suggesting gapped excitations \cite{quintero-castro}. In inelastic neutron scattering three gapped and dispersive magnetic excitation branches are observed with a spin gap $\Delta = 3.5$ meV, a maximum energy $E_{max} = 7.0$ meV and a comparably large dispersion of $\approx 3.5$ meV. A very satisfactory theoretical modeling is found for an intradimer exchange coupling $J_0 = 5.55$ meV and comparably large interdimer couplings summing to $J' \approx 3.58$ meV \cite{quintero-castro}. In ESR an excitation found at 1.250 Thz ($\mathrel{\widehat{=}} 5.17$ meV) agrees well with that energy scale \cite{deisenhofer}. Its linewidth shows a strong broadening to higher temperatures due to a relaxation process involving the higher energy $d_{x^2-y^2}$ orbital state. The resolved orbital gap $\Delta_{orbital} = 388$ K is comparably small.

In high magnetic fields Sr$_3$Cr$_2$O$_8$ shows Bose-Einstein condensation of diluted magnons. The magnetization increases with a threshold field at 30.4 Tesla and a saturation at 62 Tesla, respectively \cite{aczel}. Summarizing the available experimental data, Sr$_3$Cr$_2$O$_8$ demonstrates a hierarchy of energy and temperature scales given by the orbital gap, the Jahn-Teller transition $T_S$, the magnetic exchange interaction and the spin gap. Most noteworthy is the small separation of thermal energy to $\Delta_{orbital}$ that still allows a considerable occupation of higher lying orbital states for $T < T_S$.

Here we report on phonon anomalies, quasielastic scattering as well as magnetic Raman scattering in single crystals of Sr$_3$Cr$_2$O$_8$ that reveal an extended fluctuation regime below the Jahn-Teller transition, $T_S$. We attribute these dynamics to the interplay of lattice and electronic degrees of freedom and a comparably small orbital gap.

\section{Experimental details}

Single crystals of Sr$_3$Cr$_2$O$_8$ were grown by a floating zone
method as described previously \cite{islam}. The resulting crystal size was about $1 \times 0.5
\times 0.5$ mm$^3$. The crystals have been fully characterized with respect to their structural
and magnetic properties using X-Ray scattering, magnetic susceptibility, neutron scattering \cite{quintero-castro}, specific heat, and infrared absorption \cite{deisenhofer}.

Raman scattering experiments were performed in quasi-backscattering geometry, using a  $\lambda =
532$ nm solid state laser, a $\lambda = 561$ nm solid state laser and an Ar-Kr ion laser at
$\lambda = 488$ nm. The laser power was set to 7.5 mW with a spot diameter of approximately 100
$\mu$m to avoid heating effects. All measurements were carried out in an evacuated cryostat in the temperature range from 3 K to 360 K. The spectra were collected via a triple spectrometer
(Dilor-XY-500) by a liquid nitrogen cooled CCD (Horiba Jobin Yvon, Spectrum One CCD-3000V).

\section{Experimental Results}

\subsection{Phonons and quasielastic scattering}
For temperatures above $T_S$, the factor group analysis for the hexagonal space group
$R\overline{3}m$ yields 11 Raman-active phonon modes: $\Gamma_{Raman} = 5 \cdot$A$_{1g} + 6 \cdot$E$_g$.
Below $T_S$ the structure changes to the space group $C2/c$, which yields a total of 39
Raman-active phonons: $\Gamma_{Raman} = 19 \cdot$A$_g + 20 \cdot$B$_g$. The corresponding Raman
tensors are given by:

\begin{center}

\begin{widetext} \mbox{A$_{1g}$=$\begin{pmatrix} a & 0 & 0\\ 0 & a & 0\\ 0 & 0 & b\\
\end{pmatrix}$

, E$_{g,1}$=$\begin{pmatrix} c & 0 & 0\\ 0 & -c & d\\ 0 & d & 0\\ \end{pmatrix}$

, E$_{g,2}$=$\begin{pmatrix} 0 & -c & -d\\ -c & 0 & 0\\ -d & 0 & 0\\ \end{pmatrix}$

, A$_{g}$=$\begin{pmatrix} a & d & 0\\ d & b & 0\\ 0 & 0 & c\\ \end{pmatrix}$

, B$_{g}$=$\begin{pmatrix} 0 & 0 & e\\ 0 & 0 & f\\ e & f & 0\\ \end{pmatrix}$.} \end{widetext}
\end{center}

Raman scattering experiments were performed in different light polarizations at room temperature
($T = 295$ K) and at 3 K to distinguish between these modes, as shown in Fig. 1. Here, ($zz$)
polarization denotes a polarization of the incoming and outgoing light parallel to the
crystallographic $c$ axis, i.e. the axis, along which the spin dimers are realized. In ($yy$)
polarization, both incoming and outgoing light is polarized perpendicular to the crystallographic $c$
axis and parallel to the hexagonal plane, whereas crossed polarization is used in ($yz$) configuration.

\begin{figure}
\label{figure1}
\centering
\includegraphics[width=8cm]{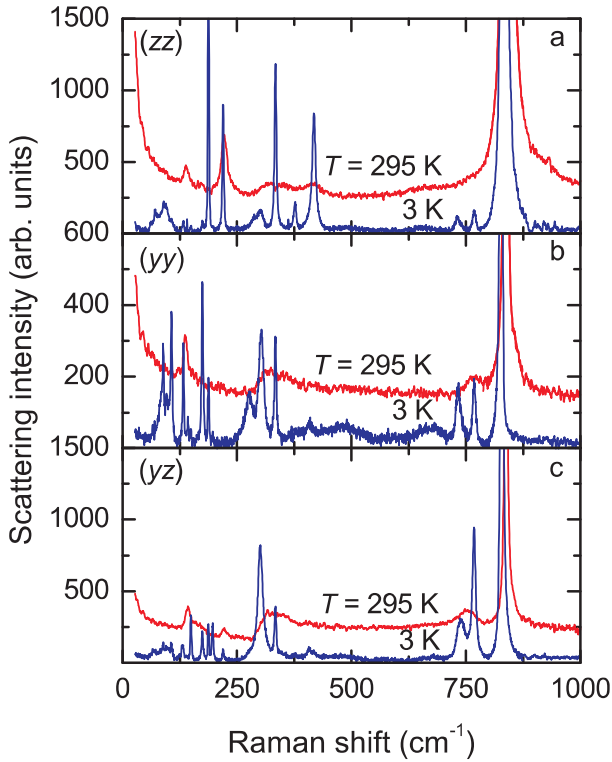}
\caption{(Colour online) Raman spectra collected at room temperature (i.e. 295 K, red curve) and
at lowest temperature (i.e. 3 K, blue curve) in different polarizations: a) ($zz$) polarization,
b) ($yy$) polarization, c) ($yz$) polarization. The depicted spectra range from 30 to 1000
cm$^{-1}$. The spectra are shifted in intensity for clarity.}
\end{figure}

At room temperature only the A$_{1g}$ phonon modes contribute to the ($zz$) and ($yy$)
polarization, while in crossed configuration only E$_g$ modes are allowed. In the structurally
distorted, low temperature phase, A$_g$ modes show up in ($zz$) and ($yy$) polarization, while
B$_g$ modes only contribute to crossed polarization.

In the spectra of Fig. 1 taken at $T = 295$ K we identify 10 out of the expected 11 phonons. With
decreasing temperatures the spectra change drastically. However, the largest changes are observed only for temperatures far below
$T_S$. At $T = 3$ K we identify a total number of 27 out of 39 expected phonon modes.
The corresponding mode frequencies and their symmetry
assignments are given in table 1. The discrepancy between the observed and expected modes in the
low temperature phase can be due to a lack of phonon intensity of particular modes as well as due
to their overlap with larger intensity excitations.

\begin{figure}
\label{figure2}
\centering
\includegraphics[width=8cm]{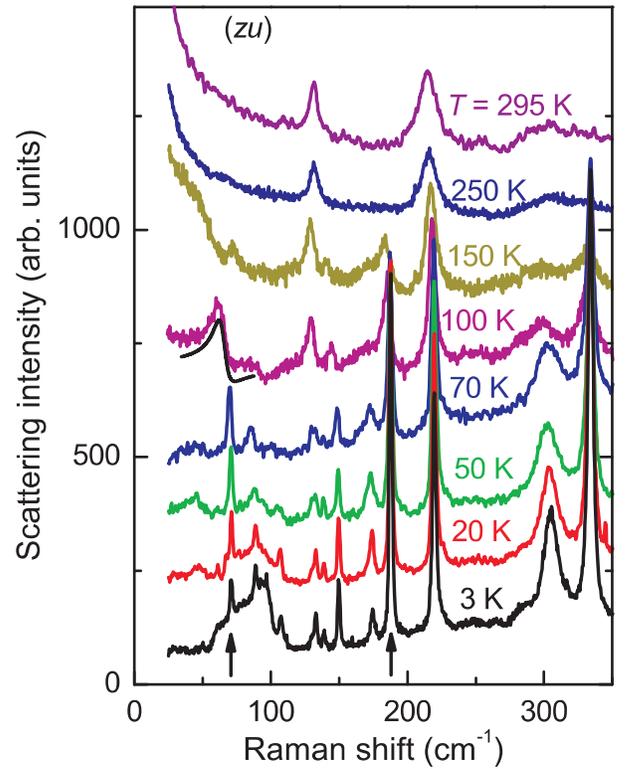}
\caption{(Color online) Temperature development of the low- and mid-frequency range (i.e. 25 - 350 cm$^{-1}$) from 3 to 295 K in ($zu$) polarization, i.e. without an analyzer; spectra are shifted in intensity for
clarity. The two arrows mark the phonons at 71 and 188 cm$^{-1}$.}
\end{figure}

In the intermediate temperature regime we observe distinct changes in the phonon spectra, i.e. decreasing phonon line width and intensity gain. Figure 2 displays the temperature evolution of the low- and
mid-frequency range of the Raman spectrum (from 25 cm$^{-1}$ to 350 cm$^{-1}$). In particular the phonons at 188 and 334 cm$^{-1}$ show a strong temperature dependence for temperatures $T < T_S$.

\begin{figure}
\label{figure3}
\centering
\includegraphics[width=8cm]{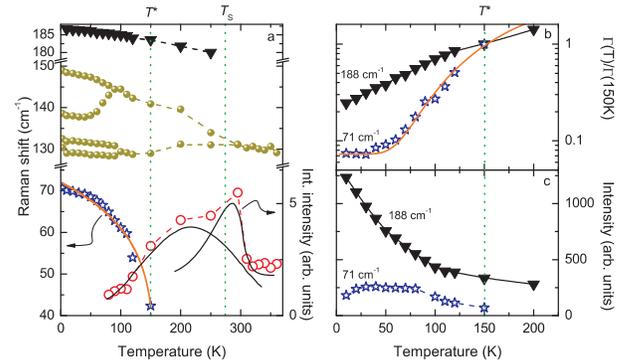}
\caption{(Color
online) a) Temperature evolution of the phonon frequency at 71 (blue stars), 129, 132, 138 and 148 (yellow
spheres), and 188 cm$^{-1}$ (black triangles). The solid orange line is a fit to the 71 cm$^{-1}$ data. The red circles display the integrated intensity of the quasielastic scattering. b) Normalized line width
$\Gamma(T)/\Gamma(150K)$ of the 188 and 71 cm$^{-1}$ phonon modes. The solid
orange line is a fit to the 71 cm$^{-1}$ data. c) Phonon intensity at 188 and 71
cm$^{-1}$. The data of the 71 cm$^{-1}$ mode has been scaled by a factor of 2; the 188 cm$^{-1}$ data has been shifted by 250 arb. units for clarity. The green dashed lines in a), b), and c) indicate $T_S = 275$ K and $T^* \approx 150$ K.}
\end{figure}

In Fig. 3 a) the frequencies of important phonons are plotted as function of temperature. For the group of phonons in the
frequency range 125 - 150 cm$^{-1}$ we observe two phonon splittings with decreasing temperature.
The first one evolves very gradually, starting around $T_S$. The second, more pronounced one evolves at lower temperatures and marks a characteristic temperature $T^* \approx 150$ K. While the phonon at 188 cm$^{-1}$ exhibits an energy shift by about 4$\%$ over the whole temperature range, the phonon at 71 cm$^{-1}$, which lies close in energy to two magnetic
modes (as described below), shows an anomalous softening ($\approx 40\%$ between 3 and 150 K). Fitting this energy to a power law, i.e. $\omega = A \cdot \left| T_C-T \right|^{\beta}$, leads to a good description in the $T^*$ regime with $A = 31.8$ cm$^{-1}$/K , $T_C = 156$ K and $\beta = 1/6 \pm 0.016$. From a mean-field approximation, one expects for a soft mode a critical exponent of $\beta = 1/2$. This is far from the fit to our data. It is noteworthy that $T_C$ of this fit is inconsistent with $T_S$ but comparable with $T^*$. This suggests that this critical behavior can be used as a second, independent way to determine the characteristic temperature $T^*$. For all phonon modes the temperature dependence is reduced for temperatures below $T = 50$ K. One could describe this as a stabilization of the lattice.

Upon approaching $T_S$, a broad, quasielastic ($E \approx 0$) scattering contribution with Lorentzian lineshape and A$_{1g}$ symmetry increases in intensity and decreases again continuously for temperatures below approximately 200 K, see Fig. 2 and the open (red) circles in Fig. 3 a) for its intensity. The data clearly reveal pronounced fluctuations and can be deconvoluted into two contributions, a sharp onset at $T_S$ and a broad maximum at lower temperatures with a softer leveling off below $T^*$, as indicated by the solid black lines. This behavior is in complete contrast to the evolution of the phonon intensity. One could compare this contribution to a so-called central mode, induced by fluctuations at a structural phase transition. However, in this case for temperatures below $T_S$ the fluctuations should be suppressed and no second maximum is expected. Therefore structural fluctuations are unlikely to be the origin for this scattering contribution. Further candidates are orbital or spin fluctuations. Such fluctuations as energy density fluctuations should scale with a corresponding contribution to the specific heat. Indeed, also in specific heat measurements a second, shallow maximum at $T = 200$ K has been observed, see inset of Fig. 3 in Ref. \cite{deisenhofer}.

In Fig. 3 b) the temperature dependence of the line width (full width at half maximum) for the phonons at 188 and 71 cm$^{-1}$ is shown. The phonon at 71 cm$^{-1}$ shows a dramatic decrease in linewidth by a factor of 11 between 150 and 3 K. In addition, there is a change in lineshape from an asymmetric Fano line to a symmetric Lorentzian. This is indicated by the black solid line fitted to the 100 K spectrum in Fig. 2. In contrast, the line at 188 cm$^{-1}$ shows only a moderate decrease in linewidth with a slight change in slope at $T^*$. We find that an activated function $\Gamma = B + C \cdot e^{-\Delta_{JT} / k_B T}$, with $B = 0.09$ cm$^{-1}$, $C = 18.1$ and $\Delta_{JT}/k_B = (450 \pm 20)$ K provides a very reasonable description to the temperature dependence of the 71 cm$^{-1}$ phonon linewidth. The characteristic energy $\Delta_{JT}$ is very similar to an orbital gap $\Delta_{orbital} = 388$ K that has been found in a previous ESR study. Here the ESR absorption line increases exponentially with increasing temperature due to an Orbach spin relaxation process \cite{deisenhofer}.
A similar relation between phonon and ESR linewidth has been observed in the Jahn-Teller system KCuF$_3$ \cite{gnezdilov, eremin}.

In Fig. 3 c) the phonon intensity is plotted as a function of temperature for the two lines at 188 and 71 cm$^{-1}$. The intensity of the 188 cm$^{-1}$ mode displays a strong increase for temperatures below $T^*$ without saturation. On the other hand, the intensity of the phonon at 71 cm$^{-1}$ saturates at $T \approx 70$ K.

\begin{figure}
\label{figure4}
\centering
\includegraphics[width=8cm]{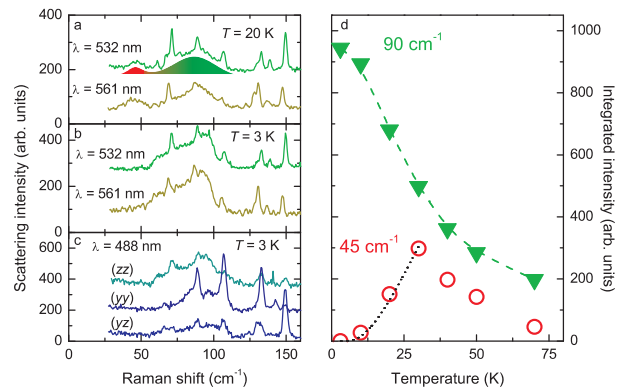}
\caption{(Color online) a) $\&$ b) low energy range of Raman spectra (ranging from 25 to 165 cm$^{-1}$) taken at
$T = 20$ K (a) and $T = 3$ K (b) with the excitation wavelengths $\lambda = 532$ nm and $\lambda = 561$ nm in ($zu$) polarization. The filled area under the spectrum in a) are fits to the
modes at 45 and 90 cm$^{-1}$. c) Low energy spectral range in different
polarizations obtained at $T = 3$ K with the $\lambda = 488$ nm laser line. d) Bose-corrected integrated intensity of the 45 and 90 cm$^{-1}$ modes. The black dotted line corresponds to a fit to the Bose-Einstein statistics of an excited level with $\Delta = 40$ K.}
\end{figure}

\subsection{Magnetic scattering}

Figure 4 a) -- c) zooms into the low energy part of the Raman
spectra. Two modes are observed at around 45 and 90 cm$^{-1}$ with a larger linewidth compared to the previously discussed phonon modes. While the 90 cm$^{-1}$ mode is most prominent at lowest temperature, the one at 45 cm$^{-1}$ exhibits a finite intensity only in the temperature interval 10 K $< T <$ 75 K. To confirm that these modes are of intrinsic nature we excite the sample with a different laser wavelength and compare the spectra as shown in panel a) and b). Fits to the two modes are shown by a highlighted area
below the $\lambda = 532$ nm spectrum in Fig. 4 a). In panel c) different polarizations at $T = 3$ K are probed with the $\lambda = 488$ nm excitation wavelength. Finally, the temperature
dependence of the Bose-corrected broad modes' integrated intensities is plotted in Figure 4 d). While the 90 cm$^{-1}$ mode increases steadily with decreasing temperature, the 45 cm$^{-1}$ mode clearly shows a maximum in intensity at around 30 K, before further decreasing and finally vanishing at lowest temperature. The intensity decrease fits well to a Bose-Einstein factor of an excited level as shown by the dotted line, $n_i \sim 1 / [e^{(E_i - \Delta )/T}-1]$, with $\Delta = 40$ K ($\mathrel{\widehat{=}} 28$ cm$^{-1}$). This is in excellent agreement with the value for the spin gap reported by neutron scattering \cite{quintero-castro}.

\begin{table*} \caption{\label{tab:table1}Phonon frequencies in cm$^{-1}$ with their respective
symmetry assignments for the two structural phases $R\overline{3}m$ (at $T = 295$ K) and $C2/c$
(at 3 K).}
\begin{ruledtabular} \begin{tabular}{ c|c||c|c|c }
 \multicolumn{2}{c}{$T = 295$ K, $R\overline{3}m$}&\multicolumn{2}{c}{$T = 3$ K, $C2/c$}& \\
 Phonon&Symmetry&Phonon&Symmetry&Comment\\
 frequency [cm$^{-1}$]&assignment&frequency [cm$^{-1}$]&assignment& \\ \hline
  & &71&A$_g$&frequency $\&$ linewidth anomalies\\
  & &88&A$_g$\\
  & &106&A$_g$\\
  & &131&B$_g$&splits out of 132 cm$^{-1}$ mode\\
  132&A$_{1g}$ + E$_g$&133&A$_g$&splits out of 132 cm$^{-1}$ mode\\
  & &139&A$_g$&splits out of 132 cm$^{-1}$ mode\\
  & &149&B$_g$&splits out of 132 cm$^{-1}$ mode\\
  & &175&B$_g$\\
  & &188&A$_g$&strong increase in intensity\\
  & &197&B$_g$\\
  222&A$_{1g}$&220&A$_g$\\
  & &276&A$_g$\\
  & &288&A$_g$\\
  300&E$_g$&302&B$_g$&large linewidth\\
  330&E$_g$&334&A$_g$ + B$_g$&strong increase in intensity\\
  & &377&A$_g$\\
  & &407&B$_g$\\
  415&A$_{1g}$&418&A$_g$\\
  & &732&A$_g$\\
  750&E$_g$&740&B$_g$\\
  765&A$_{1g}$&768&A$_g$ + B$_g$\\
  & &824&B$_g$\\
  840&A$_{1g}$ + E$_g$&829&A$_g$ + B$_g$\\
  & &835&A$_g$\\
\end{tabular} \end{ruledtabular} \end{table*}

\section{Discussion}

We will first focus on the temperature evolution of the phonon modes before discussing magnetic and orbital Raman scattering.

The present experimental data allow to pinpoint a few experimental facts about Sr$_3$Cr$_2$O$_8$ as given in the introduction. However, important information about the evolution of electronic and structural correlations is missing as no detailed temperature studies are available up to now. Based on neutron diffraction \cite{chapon} and specific heat studies \cite{deisenhofer} the Jahn-Teller transition exists around $T_S = 275-285$ K. Infrared spectroscopy revealed the appearence of new polar phonon modes below this transition \cite{deisenhofer}. On the basis of the CEF level scheme of Cr$^{5+}$ in tetrahedral coordination it is very probable that an order-disorder transition of the orbital occupation number dominates this transition. Whether the electronic and structural correlations develop concomitant and continuously or discontinuously has to be further evaluated. Here Raman scattering data can give decisive information as not only symmetry information but also electronic polarizabilities of the respective states are probed. Raman scattering is a fast technique that may also detect local distortions as it relies on an optically induced virtual electron-hole pair. There also exists the chance to observe collective orbital excitations. However, here much less experimental knowledge exists.

An important observation is the nonexistence of sudden changes of the Raman phonon spectra between 295 K and 250 K, see Fig. 2. This means that the coherent part of lattice distortions develops only at lower temperatures. In contrast, the quasielastic scattering, a probe of fluctuations with similar evolution as the specific heat, discontinuously sets in for $T = T_S$. We therefore conclude an extended orbital fluctuation regime from $T_S$ down to approximately $T^* = 150$ K. For $T < T^*$ the phonon modes show a gradual increase of intensity and a rapid increase below approximately 110 K. Below also 110 K distinct phonon splittings show up and the quasielastic scattering is suppressed. Therefore $T^*$ marks a crossover with a suppression of orbital fluctuations and an onset of long range structural distortions.

The observed dynamics is different from quasi-one-dimensional correlated electronic systems like BaVS$_3$ \cite{choi-09} or weakly doped manganites where persisting orbital fluctuations are attributed to
an entangled instability of the orbital, electronic, and lattice subsystems \cite{tokura}. Since Sr$_3$Cr$_2$O$_8$ has no electronic instability or low dimensionality, the origin of the fluctuations should be sought in the relation between the orbital ordering and the structural instability. Introducing a ratio of the Jahn-Teller gap versus the structural phase transition temperature, $\Delta_{JT}/T_S = 1.4$ -- 1.6, shows that thermal fluctuations even well below $T_S$ can smear out the Jahn-Teller gap and thus destabilize orbital ordering.

We carefully inspected whether collective orbital excitations (orbitons) show up in the frequency range of $\Delta_{JT} \approx$ 270 -- 330 cm$^{-1}$. At room temperature we observe a highly-damped peak in the respective energy interval (see Fig. 2). With lowering temperature two modes develop at about 300 and 330 cm$^{-1}$. In contrast to other sharp peaks at 3 K, the mode at 300 cm$^{-1}$ ($\mathrel{\widehat{=}} 430$ K) is relatively broad. This indicates the possibility that the corresponding mode is an orbital excitation mixed with a phonon. A similar observation was reported for the vanadates $R$VO$_3$, where orbital excitations couple to the Jahn-Teller phonon \cite{miyasaka}. This phenomenology is different from observations in the three dimensional perovskites LaTiO$_3$ and YTiO$_3$ as in these compounds the energy scale of the orbitons (2700 K) is clearly separated from the phonon system \cite{ulrich}.

Further evidence for mixing of phonon and orbital degrees of freedom is gained from the temperature dependence of the 71 cm$^{-1}$ phonon mode, in energy, linewidth and lineshape.
Actually, this dependence shares similarities with a crystalline electric field transition in the rare earth compound Pr$_2$CuO$_4$ \cite{sanjurjo}. However, in a $3d$ electron system such modes should be broadened by dispersion. In band structure calculations \cite{radtke} for Sr$_3$Cr$_2$O$_8$ the respective energy scale is of the order of 300 meV, i.e. also one order larger than $\Delta_{JT}$. Therefore the origin of the 71 cm$^{-1}$ mode is very likely a mixed phonon-orbiton mode.

Next, we will focus on the magnetic excitations in Sr$_3$Cr$_2$O$_8$. As reported previously \cite{quintero-castro}, these excitations form a gapped magnon with a gap size of $\Delta = 28$
cm$^{-1}$. In Fig. 4 a), two magnetic modes are observed at low, but finite temperatures. The high energy magnetic mode at 90 cm$^{-1}$ is observed in all measured polarizations, i.e. ($zu$),
($zz$), ($yy$) and ($yz$), but has its strongest contribution in ($zz$) and ($zu$) polarization,
i.e. the polarization along the dimer direction, while its intensity is weakest in crossed, i.e.
($yz$) polarization. Since the magnetic Raman scattering intensity is given by $I \propto
\left| \langle i \right| \sum_{i,j} \vec S_i \cdot \vec S_j \left| j \rangle \right|^2$, it scales in each polarization with the strength of exchange interactions between dimers. In this sense, the presence of the substantial spectral weight in the ($yy$) polarization implies moderate interdimer interactions in the $ab$ plane.

The onset energy of the magnetic excitations is roughly 56 cm$^{-1}$, which corresponds to 7 meV or 81 K, and is therefore twice the size of the spin gap. The high energy cut-off at 118 cm$^{-1}$ corresponds to 14.75 meV. As the triplet branch of excitations observed in inelastic neutron spectroscopy reaches up to 7.2 meV, we attribute this mode to two-magnon scattering, i.e. a scattering process involving the simultaneous spin flip of two neighboring spins, hence costing twice the magnetic exchange energy. The intensity of the two-magnon mode is plotted over
temperature in Fig. 4 d) (green triangles). It continuously decreases in intensity as the temperature is increased and fully depletes below 100 K. This is associated with a thermal population of the singlet ground state into an excited triplet state.

The intensity of the mode at 45 cm$^{-1}$, on the other hand, shows a clear maximum at around 30 K before steadily decreasing towards $\approx 80$ K. Therefore, we can rule out one-magnon
scattering, which would show a monotonous increase in intensity down to lowest temperature.
Instead, a three-magnon scattering process might account for this observed mode
\cite{els-cugeo3}. Three-magnon scattering can occur through a transition from one singlet and one excited triplet state to two neighboring triplet states, i.e. $\left(\mid \uparrow \downarrow
\rangle - \mid \downarrow \uparrow \rangle \right) \otimes \left( \mid \uparrow \downarrow \rangle
+ \mid \downarrow \uparrow \rangle \right) \rightarrow \mid \uparrow \uparrow \rangle \otimes \mid
\downarrow \downarrow \rangle$. Therefore it can only arise in the environment of thermally
excited triplet states at finite temperatures. Such a thermal population is described by
the Bose-Einstein statistics of the excited level. Our data is in good agreement with theory, as the fit to the integrated
intensity at low temperatures in Fig. 4 d) shows. Upon additional temperature increase, one
furthermore increases the population of triplets, which eventually leads to saturation. This
causes the intensity of the three-magnon mode to drop again towards high temperatures. The onset
of the mode is at roughly 30 cm$^{-1}$. As a three-magnon process corresponds to a transition from
the triplet branch at the $\Gamma$-point of the Brillouin zone to the two-magnon band, the onset
energy can be estimated by $E_{2M,onset} - \Delta \approx \Delta = 28$ cm$^{-1}$ which fits well
to the observed value. The cut-off energy would therefore be expected around $E_{2M,cutoff} -
\Delta \approx 90$ cm$^{-1}$. However, it can only unambiguously be traced up to 56 cm$^{-1}$,
i.e. the onset of the two-magnon signal. Above this energy the two modes overlap, therefore a
clear determination of the cut-off energy is rather difficult.

Remarkably, such a three-magnon scattering has also been reported in the weakly coupled spin dimer compound KCuCl$_3$ with a spin gap of $\Delta = 31$ K \cite{choi-05} and in the Spin-Peierls system CuGeO$_3$ with $\Delta = 23$ K \cite{els-cugeo3}. All three compounds show a maximum in intensity around the spin gap temperature.
However, for KCuCl$_3$ the three-magnon signal is observed up to $T \approx 5 \cdot \Delta$, while for Sr$_3$Cr$_2$O$_8$ it is obscured by the lattice fluctuations even at $T \approx 2 \cdot \Delta$.

All three systems show an instability in large magnetic fields. For KCuCl$_3$ and Sr$_3$Cr$_2$O$_8$ this is a Bose-Einstein condensation of diluted magnons, while CuGeO$_3$ becomes an incommensurate antiferromagnet. The common observation of three-magnon Raman scattering and a field-induced transition is due to enhanced hopping, i.e., a not too large gap.

\section{Summary}
In conclusion, our Raman data reveals an extended temperature regime in which orbital fluctuations dominate the lattice dynamics of Sr$_3$Cr$_2$O$_8$. The energy scale of the orbital ordering, driven by a Jahn-Teller distortion, is estimated to be $\Delta_{JT} = 450$ K in good agreement with ESR. This energy is also comparable to typical phonon frequencies of the tetrahedral Cr-O coordinations and thermal energies for $T < T_S = 275$ K. This promotes an intricate interplay between orbital-lattice and spin-orbital coupled quantities. In addition, the analysis of two- and three-magnon scattering processes in the orbitally ordered state allows us to determine the spin gap $\Delta = 40$ K.

\begin{acknowledgments}
We acknowledge valuable discussions with Yu. G. Pashkevich and support from DFG, B-IGSM and NTH. KYC acknowledges financial support from Humboldt foundation and Korea NRF Grant No. 2009-0093817. JD acknowledges support from TRR80 (Augsburg-Munich).
\end{acknowledgments}

\end{document}